\documentclass[%
reprint,
superscriptaddress,
amsmath,amssymb,
aps,
pra,
floatfix,
]{revtex4-1}

\usepackage{graphicx}
\usepackage{dcolumn}
\usepackage{bm, bbm}
\usepackage{xcolor}
\usepackage{float}
\usepackage{comment}

\newcommand{\ket}{{\rangle}}
\newcommand{\bra}{{\langle}}

\begin{document}

\preprint{PAE V3}

\title{Parameterized Attenuated Exchange for Generalized TDHF@$v_W$ Applications}

\author{Barry Y. Li}
\affiliation{Department of Chemistry and Biochemistry, University of California, Los Angeles, Los Angeles, CA, 90095, USA}
\author{Tim Duong}
\affiliation{Department of Chemistry and Biochemistry, University of California, Los Angeles, Los Angeles, CA, 90095, USA}
\author{Tucker Allen}
\affiliation{Department of Chemistry and Biochemistry, University of California, Los Angeles, Los Angeles, CA, 90095, USA}
\author{Nadine C. Bradbury}
\affiliation{Department of Chemistry, Princeton University, Princeton, NJ, 08544, USA}
\author{Justin R. Caram}
\email{jcaram@chem.ucla.edu}
\affiliation{Department of Chemistry and Biochemistry, University of California, Los Angeles, Los Angeles, CA, 90095, USA}
\author{Daniel Neuhauser} 
\email{dxn@ucla.edu}
\affiliation{Department of Chemistry and Biochemistry, University of California, Los Angeles, Los Angeles, CA, 90095, USA}

\date{\today}

\begin{abstract}
Building upon our previously developed time-dependent Hartree-Fock (TDHF)@$v_W$ method, based on many-body perturbation theory and specifically the Bethe-Salpeter Equation (BSE), we introduce a parameterization scheme for the attenuated exchange kernel, $v_W(|r - r'|)$. In the original method, $v_W$ was determined individually for each system via an efficient stochastic short-time TD Hartree propagation for the screened Coulomb interaction, $W(r,r')$. The new parameterization leverages {photophysical} similarities in exciton binding energies (or exchange interaction attenuation) among molecules with comparable static dielectric responses. We parameterize the inverse dielectric function using a low-order polynomial with error function apodization, calibrated on a few representative molecules, each with its own $v_W$. Using only 7 parameters, the parameterized $v_W$ is fully grid-independent and broadly applicable within a family of molecules. This enables TDHF@$v_W$ that retains BSE-level accuracy, achieving a mean absolute error of $\sim0.1$ eV compared to experimental optical gaps and representing a five- to ten-fold improvement over conventional TD density functional theory or TDHF while reducing the cost to that of standard TDHF.
\end{abstract}

\maketitle

\section{Introduction}
The GW-Bethe-Salpeter equation (GW-BSE) approach has become a very popular method to accurately calculate the optical absorption spectra of molecular systems. Within the framework of many-body perturbation theory (MBPT), the success of the method lies in the explicit inclusion of an effective screened Coulomb interaction kernel, $W$.\cite{Boulanger2014,Jacquemin2017} An improved description of electron correlation through $W$ makes the GW-BSE method capable of capturing the complex multi-configurational character of excited states, such as those present in delocalized, highly conjugated molecular systems with closely spaced energy levels that are not well described by time-dependent density functional theory (TDDFT).\cite{LeGuennic2015,Jacquemin2007,neuhauser2015} 

Employing the static and Tamm-Dancoff approximations, the BSE describes couplings of singlet electron-hole pairs, i.e., excitons, through the resonant matrix, $A$: 
\begin{equation}
\label{TDA}
    {{A_{ia,jb}}} = (\varepsilon_a-\varepsilon_i+\Delta)\delta_{ij}\delta_{ab}+2(ia|jb)-(ab|W|ij),
\end{equation}
where the $N_vN_c\times N_vN_c$ valence-conduction product basis is composed of a generalized Kohn-Sham DFT eigensystem. Indices $i,j,...$ refer to valence (hole) and $a,b,...$ to conduction (electron) states. A GW-derived scissor energy correction, $\Delta$, is then applied to the independent-particle term of $A$ to include single-particle self-energy effects.\cite{bradbury_optimized_2023,Allen2024} Assuming real orbitals, the bare Coulomb integrals are:
\begin{equation}
    (ia|jb)=\int dr \, dr' \, \phi_i(r) \phi_a(r) |r-r'|^{-1} \phi_j(r') \phi_b(r'),
\end{equation}
and the screened direct interaction matrix elements are:
\begin{equation}
\label{Wmat}
    (ab|W|ij) = \int dr \, dr' \, \phi_a(r) \phi_b(r) W(r, r') \phi_i(r') \phi_j(r').
\end{equation}
Replacing the effective interaction, $W(r,r')$, with the bare Coulomb potential, $|r-r'|^{-1}$, and letting 
$\Delta=0$ {(i.e., not requiring a separate GW calculation on the HOMO and LUMO levels)} converts Eq. \ref{TDA} to the Casida formulation of time-dependent Hartree-Fock (TDHF).\cite{Negele1982} {Following the frequency-domain approach of Ref. \cite{Sereda2024}, this is equivalent to the random phase approximation with exchange (RPA-X) of quantum chemistry.}

Constructing and storing the matrix elements of Eq. \ref{Wmat} becomes the major computational bottleneck of a BSE calculation with an overall scaling of $\mathcal{O}(N^4)$.  It is standard practice to obtain $W(r,r')$ within the RPA.\cite{Louie_BSE_2000} In the frequency domain, this requires evaluating the static limit of the full dielectric matrix, $\epsilon(r,r'')$, whose inverse relates to $W$ by:
\begin{equation}
    W(r,r') = \int dr'' \epsilon^{-1}(r,r'')|r''-r'|^{-1}.
\end{equation}
Meanwhile, in the time-domain, the elements of Eq. \ref{Wmat} are obtained through a time-dependent Hartree (TDH) propagation by perturbing and propagating all occupied states.\cite{Vlek20181}  

We have shown in several previous works that efficient stochastic techniques can enable GW-BSE calculations for systems consisting of hundreds to thousands of valence electrons. This includes an exact division of $W$ to simpler terms $W\equiv v_W+\{W-v_W\}$, where $v_W(|r-r'|)$ is a momentum-space diagonal (i.e., translationally invariant) attenuated exchange kernel that captures the bulk of the effect of $W$, and $\{W-v_W\}$ is the remaining difference that is stochastically sampled.\cite{bradbury_bethesalpeter_2022,bradbury_neargap_2023,bradbury_optimized_2023,Allen2024}

The methodology is detailed in Refs. \cite{bradbury_optimized_2023,Sereda2024,no_more_gap} and summarized here. We wish to approximate the effect of $W(r,r')$ on occupied-occupied pair densities by minimizing the objective $\sum_{ij}(ij|(W-v_W)^2|ij)$ to obtain a simpler screened exchange kernel, $v_W$. First, a set of statistically independent stochastic occupied orbitals is introduced, $\bar{\beta}(r) = \sum_i(\pm1)\phi_i(r)$ and $\bar{\bar{\beta}}(r) = \sum_i(\pm1)\phi_i(r)$, which yield a pair density $\beta(r)\equiv \bar{\beta}(r)\bar{\bar{\beta}}(r)$. With a few manipulations, the optimal form of $v_W$ is expressed in reciprocal space: 
\begin{equation}
   \label{eq:vw_opt}
   v_W(k) =
   \Big\{ \frac{\beta^*(k)\bra k |W|\beta\ket  }
   { |\bra k |\beta \ket |^2}\Big\}_{\beta} ,
\end{equation}
where $\{...\}$ indicates a statistical average over the number of stochastic samples, $\beta$. The action of the many-body $W$ on a random pair density, $W_\beta(k)\equiv\bra k |W|\beta\ket$, is obtained in real-space through a stochastic TDH propagation with a source potential derived from $\beta(r)$. As verified in our previous studies, typically 2000 stochastic samples are sufficient for convergence of both $v_W$ and the BSE absorption spectrum.\cite{no_more_gap,bradbury_optimized_2023} 

\begin{figure}[H]
\centering
\includegraphics[width=3.5in]{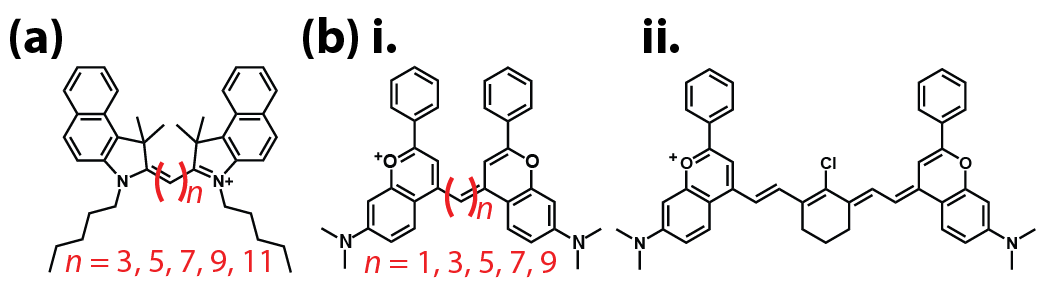}
\caption{\label{fig:fig_1} \textbf{(a)} ICG dye structures with various polymethine chain lengths. \textbf{(b)} \textbf{i.} Flav dye structures, where $n=7$ means LFlav-7 (the linear Flav-7); \textbf{ii.} Flav-7.\cite{no_more_gap}}
\end{figure}

The TDHF@$v_W$ method is obtained by replacing $W$ with $v_W$ for the electron-hole interaction kernel, omitting the difference, $\{W-v_W\}$.\cite{bradbury_optimized_2023} In our recent work, we applied this method to several $\pi$-conjugated, near- and shortwave-infrared dyes, including the flavylium (Flav) and indocyanine green (ICG) families of polymethine cyanine dyes (molecular structures shown in Fig. \ref{fig:fig_1}), and demonstrated good agreement between the calculated spectra and experimental measurements.\cite{no_more_gap}

\begin{figure}[H]
\centering
\includegraphics[width=3.5 in]{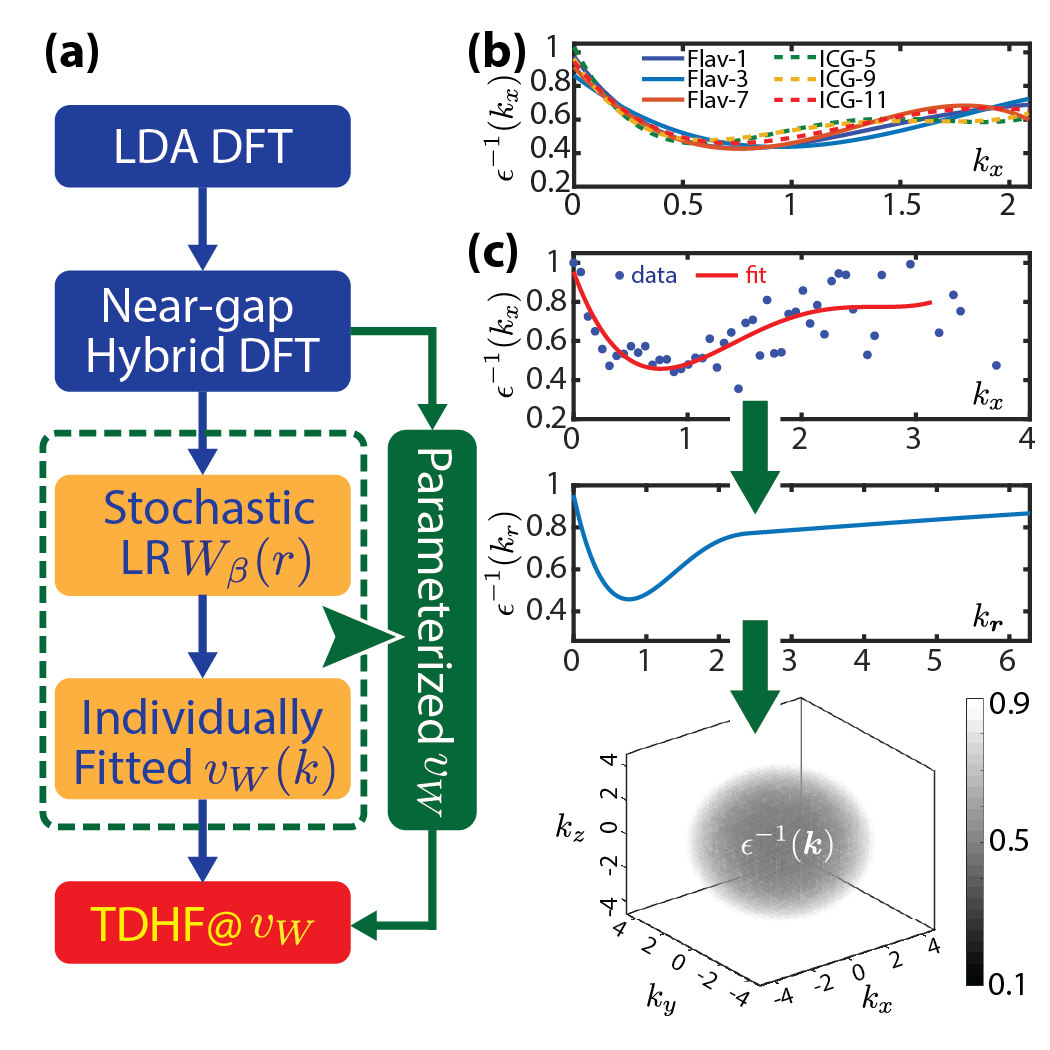}
\caption{\label{fig:fig_2} \textbf{(a)} The schematic flow of the overall computation. The parameterization (in green) of $v_W$ in this work replaces the needs of individual stochastic linear-response $W_\beta$ and $v_W$ generation. \textbf{(b)} $\epsilon^{-1}(k)$ for a few Flav and ICG dyes.\cite{no_more_gap} \textbf{(c)} Schematic $\epsilon^{-1}(k)$ parameterization and regeneration process.}
\end{figure}

The computational scheme was detailed in our previous publication, Ref. \cite{no_more_gap}. Fig. \ref{fig:fig_2}a summarizes the procedure: the (semi)local density approximation (LDA) DFT is performed first, followed by a near-gap hybrid treatment to include explicit exchange.\cite{bradbury_neargap_2023,Sereda2024} Next, a small number of stochastic actions, $W_\beta$, yield the static RPA response,\cite{bradbury_bethesalpeter_2022} where $v_W$ is then individually fitted by averaging over $W_\beta$, per Eq. (\ref{eq:vw_opt}).\cite{bradbury_optimized_2023,no_more_gap} Finally, the $v_W$ is used as an attenuated exchange kernel in the TDHF calculations for optical spectra. 

We observed that the individually fitted $v_W$ are similar among these dyes (Fig. \ref{fig:fig_2}b), indicating that there exists a generalized, parameterizable form of the exchange kernel for families of molecules with similar dielectric screening. Thus, we can bypass the need of $W_\beta$ and the individually fitted $v_W$ as indicated by the green path (Fig. \ref{fig:fig_2}a). Reducing the computational cost to conventional TDHF while maintaining BSE-quality results is a long-standing goal.\cite{Sun2020} Here, we introduce a parameterization scheme (Fig. \ref{fig:fig_2}c) of $v_W$, allowing for the use of a single kernel at the TDHF level for various molecules regardless of the system or computational grid. We also provide a simple functional form of $v_W$ for families of molecules, avoiding the need to prepare the stochastic actions $W_\beta$ altogether. 

The following section describes our parameterization procedure and demonstrates its application through 3 sets of parameterized results for different molecular families: 11 polymethine cyanine dyes, 6 planar aromatic hydrocarbons, and 3 curved aromatic hydrocarbons (structures in Fig. \ref{fig:fig_1} and Appendix A). All molecular geometries are optimized under vacuum using ORCA 6.0 at the B3LYP/def2-TZVPPD level.\cite{Neese2020,Neese2022} {All geometries were optimized with ORCA’s B3LYP (distinct from Gaussian’s B3LYP/G) for consistency with Ref. \cite{no_more_gap}.} With our generalized, parameterized, functional form of $v_W$, the TDHF@$v_W$ formalism can be made even cheaper with a balance between computational complexity (at the cost of TDHF) and accuracy (at the near-BSE quality) for optical excitation spectra. By doing so, we unlock the full potential to elucidate {photophysical} properties for various classes of chromophores, improving their application in fluorescence detection techniques and the study of larger biomolecules through a simple, generalized parameterization of small molecules.

\section{Methodology}
We introduce a parameterization of $v_W(k)$ for families of molecules sharing similar static dielectric properties. The inverse dielectric function can be written as
\begin{equation}
    \epsilon^{-1}(k) = \frac{v_W}{v_{\text{bare}}}  + 1,
\end{equation}
where $v_{\text{bare}}=|r-r'|^{-1}$. Our previous article presents $\epsilon^{-1}(k)$ for multiple Flav and ICG dyes,{\cite{no_more_gap}} where we realize that they reveal a striking similarity. This suggests the possibility of a generalized $v_W$.

Starting with a small- to mid-sized dye or hydrocarbon and performing the original stochastic fitting to extract $v_W(k)$, we proceed with a one-dimensional (1D) functional fitting of the inverse dielectric function, {$\epsilon^{-1}(k_x)$}, using a $4^{\text{th}}$-order polynomial,
\begin{equation}
    f_1^x(k) = 1 + \sum_{n=0}^{4} c_n^x k^n.
\end{equation}
We fit {$\epsilon^{-1}(k_y)$} and {$\epsilon^{-1}(k_z)$} with the same procedure{, where $k_x$, $k_y$, and $k_z$ are slices along the respective $(1,0,0)$, $(0,1,0)$, and $(0,0,1)$ directions.} To get an isotropic-averaged set of polynomial coefficients,
\begin{equation}
    c_n=\frac{1}{3}\left(c_n^x+c_n^y+c_n^z\right),
\end{equation}
that are used to calculate an overall $f_1$.

We then introduce an error function tail to handle high-$k$ stochastic noise present due to the Martyna-Tuckerman technique used to avoid grid-reflection effects when generating the stochastic actions, $W_\beta$.\cite{MartynaTuckerman1999} We define a high-$k$ noise cutoff, $k_{\text{mt}}$, and the corresponding fitted function value, $f_{\text{mt}} = f_1(k_{\text{mt}})$. For $k\leq k_{\text{mt}}$, $f_1(k)$ is used. For $k > k_{\text{mt}}$,
\begin{equation}
\label{fit2}
    f_2(k) = \left( 2 - 2 f_{\text{mt}} \right)\left\{ \frac{1}{2} \text{erf} \left[ \gamma(k - k_{\text{mt}}) \right] + \frac{1}{2}\right\} + 2 f_{\text{mt}} - 1
\end{equation}
is used. Thus, the final fitted $\epsilon^{-1}(k)$ is
\begin{equation}
    \epsilon^{-1}(k) = \begin{cases}
f_1(k) & \text{for } k \leq k_{\text{mt}}, \\
f_2(k) & \text{for } k > k_{\text{mt}}.
\end{cases}
\end{equation}
{We have verified with the pyrene molecule that results are mostly insensitive to the momentum cutoff parameter $k_{\text{mt}}$ (see Appendix B): varying $k_{\text{mt}}$ from 1.0 to 1.5 Bohr$^{-1}$ (real-space $1.66–1.11 \text{\AA}$) shifts the singlet-excitation ($S_1$) peak by less than 4 meV.} The 1D parameterization results in seven parameters: five polynomial coefficients, $c_n$, the noise cutoff, $k_{\text{mt}}$, and the steepness of the error function tail, $\gamma$. Now, everything is made off-grid and easy to store, modify, and share. It is straightforward to place the parameterized 1D function back on any customizable three-dimensional grid for application via a central-symmetric extrapolation of $\epsilon^{-1}(k)$ onto $\epsilon^{-1}(k_r)$, where $k_r = \sqrt{k_x^2 + k_y^2 + k_z^2}$. 

\begin{figure*}
\centering
\includegraphics[width=6.5 in]{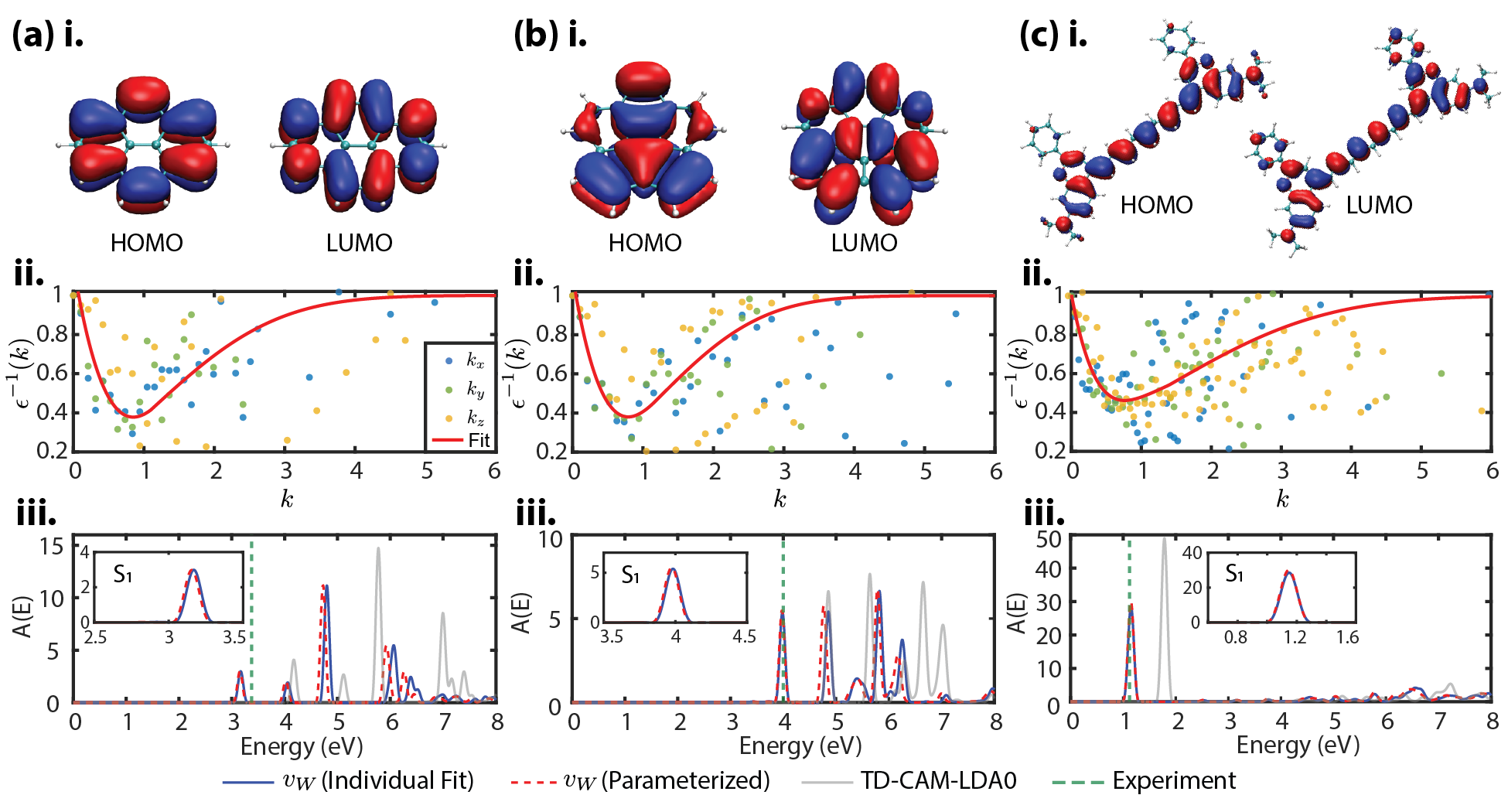}
\caption{\label{fig:fig_3} For \textbf{(a)} pyrene, \textbf{(b)} corannulene, and \textbf{(c)} Flav-9, we show \textbf{i.} the HOMO and LUMO densities calculated from near-gap hybrid DFT via the CAM-LDA0 functional; \textbf{ii} $\epsilon^{-1}(k)$ along {$(1,0,0)$, $(0,1,0)$, and $(0,0,1)$ directions} (dots), $k$ is in unit of Bohr$^{-1}$, and the parameterization is in red lines; \textbf{iii.} optical spectra calculated against the experimental $S_1$ energy via the individually fitted $v_W$, parameterized $v_W$, and TDDFT@CAM-LDA0. The experimental data are obtained from Refs. \cite{pyrene,c20,Cosco2017_ange}.}
\end{figure*}

\section{Results and Discussion}
Plane-wave (PW) pseudopotential LDA-DFT simulations are performed for all 20 test systems, followed by a near-gap hybrid DFT calculation via CAM-LDA0 functional to include explicit exact-exchange. The computational grid and the numbers of valence ($N_v$) and conduction ($N_c$) near-gap MOs used in the near-gap hybrid DFT step for each system are tabulated in Appendix C. Uniform grids are used for all the calculations, where $dx=dy=dz\sim0.5$ Bohr with verified convergence in previous publications. For both LDA and near-gap, we require the self-consistent field (SCF) energy to converge to $10^{-8}$ eV. {For convergence of the $S_1$ excitation energies}, we include $\sim5$ times more unoccupied states than occupied states in the LDA-DFT stage. For exchange energy convergence, we use all occupied MOs and $\sim3$ times more unoccupied MOs ($N_c\approx3N_v=3N_\text{occ}$) in the near-gap step. {Previous work \cite{mckeon_optimally_2022} has shown that a hybrid-DFT eigensystem can serve as an excellent starting point for excited-state calculations. We chose the CAM-LDA0 functional as it is well-suited for $\pi$-conjugated systems with a balanced treatment of short- and long-range exchange,\cite{Yanai2004} which effectively captures mid-range screening, a critical factor in exciton binding. As demonstrated in Ref. \cite{no_more_gap}, this approach yields excitation energies in good agreement with experimental results.}

All optical absorption spectra are calculated with an iterative Chebyshev solver, as in Refs. \cite{bradbury_bethesalpeter_2022,bradbury_optimized_2023,Sereda2024,no_more_gap}. The spectral widths of the absorption lines are determined by the number of polynomials used in the Chebyshev expansion, which is fixed at $3000$ terms for all simulations. We go beyond the Tamm-Dancoff approximation for all spectral calculations, including the resonant-antiresonant coupling effects between positive- and negative-frequency transitions.\cite{RARC_PRB_2009} This is at minimal additional cost as we implement sparse-stochastic sampling of the exchange kernel matrix elements. 

Three reference molecules that represent families of molecules are chosen to generate the individually fitted $v_W$ exchange kernel: pyrene for planar hydrocarbons, corannulene (C$_{20}$H$_{10}$) for curved hydrocarbons, and Flav-9 for polymethine dyes (ICG and Flav). The reference choice can be arbitrary, but we select these three due to their moderate size, good geometry convergence. Fig. \ref{fig:fig_3} shows the resulting spectra for these 3 references. The HOMO and LUMO densities are obtained from the near-gap DFT calculation using the CAM-LDA0 functional.\cite{Mosquera2016,Yanai2004,Tawada2004,no_more_gap} As expected, they are all $\pi$ and $\pi^*$ electronic structures and contribute predominately to the lowest singlet ($S_1$) excitation. Such delocalized frontier MOs require sophisticated treatment of non-local exchange over all ranges. We verified in previous work that the mid-range screening plays a crucial role in these systems for capturing correct dielectric response and exciton bindings.\cite{no_more_gap} 

\begin{table}[H]
\centering
\begin{ruledtabular}
\begin{tabular}{cccc}
 & \textbf{Pyrene} & \textbf{Corannulene} & \textbf{Flav-9} \\ \hline
$c_0$          &$\;\;\,0.06$   &$\;\;\,0.09$   &$\;\;\,0.24$  \\
$c_1$          &$-0.63$        &$-0.76$        &$-1.23$       \\
$c_2$          &$\;\;\,2.00$   &$\;\;\,2.19$   &$\;\;\,2.39$  \\
$c_3$          &$-2.20$        &$-2.21$        &$-1.93$       \\
$c_4$          &$\;\;\,0.15$   &$\;\;\,0.10$   &$\;\;\,0.01$  \\
$k_\text{mt}$  &$\;\;\,1.10$   &$\;\;\,1.20$   &$\;\;\,1.40$  \\
$\gamma$       &$\;\;\,0.50$   &$\;\;\,0.60$   &$\;\;\,0.40$  \\
\end{tabular}
\end{ruledtabular}
\caption{The resulting sets of the 7 fitted parameters of $\epsilon^{-1}(k)$ for pyrene, corannulene, and Flav-9.}
\label{tab:functional_params}
\end{table}

\begin{table*}
\centering
\begin{ruledtabular}
\begin{tabular}{>{\centering\arraybackslash}m{1.2cm} | >{\centering\arraybackslash}m{1cm} >{\centering\arraybackslash}m{1cm} >{\centering\arraybackslash}m{1cm} >{\centering\arraybackslash}m{1.22cm} >{\centering\arraybackslash}m{1cm} >{\centering\arraybackslash}m{1.74cm} >{\centering\arraybackslash}m{1cm} | >{\centering\arraybackslash}m{1.15cm} >{\centering\arraybackslash}m{1.15cm} | >{\centering\arraybackslash}m{1.8cm}}
\textbf{System} & LDA    & PBE0    & B3LYP    & LC-PBE    & $\omega$B97X & CAM-LDA0 & TDHF & $v_W$ ind. & $v_W$ par. & Expt. \\ \hline
{Flav-1}&{2.04} & {2.54} & {2.47} & {2.88}	& {2.81} & {3.29} & {2.53} & {1.73} & {1.78} & {1.90} {\cite{Cosco2017_ange}}\\
Flav-3      & 2.05 & 2.45 & 2.41 & 2.56 & 2.52 & 2.39 & 2.92 & 1.62 & 1.65 & 1.66 \cite{Cosco2017_ange}\\
Flav-5      & 1.97 & 2.26 & 2.23 & 2.25 & 2.23 & 2.06 & 2.58 & 1.34 & 1.40 & 1.50 \cite{Cosco2017_ange}\\
LFlav-7     & 1.91 & 2.12 & 2.11 & 2.04 & 2.04 & 2.01 & 2.36 & 1.35 & 1.32 & 1.26 \cite{Cosco2017_ange}\\
Flav-7      & 1.90 & 2.06 & 2.05 & 1.95 & 1.95 & 1.94 & 2.25 & 1.29 & 1.24 & 1.21 \cite{Cosco2017_ange}\\ 
Flav-9      & 1.84 & 2.00 & 2.00 & 1.86 & 1.86 & 1.84 & 2.18 & 1.15 & 1.15 & 1.12 \cite{Cosco2017_ange}\\ \hline
{\textbf{MAE}}      &{0.51}	&{0.80}	&{0.77}	&{0.81}	&{0.79}	&{1.16}	&{0.69} &{0.09} & {0.06} &  \\ \hline
%\textbf{MAE}      & 0.58 & 0.83 & 0.81 & 0.78 & 0.77 & 0.70 & 1.11 & 0.08 & 0.04 &  \\ \hline
Naph. & 4.27 & 4.57 & 4.49 & 4.93 & 4.78 & 4.62 & 5.11 & 4.23 & 4.06 & 4.10 \cite{naph_tetr1,naph_tetr2} \\
Anth. & 3.15 & 3.53 & 3.44 & 4.17 & 4.01 & 3.68 & 4.13 & 3.35 & 3.34 & 3.26  \cite{anth}\\
Tetra.   & 2.39 & 2.77 & 2.69 & 3.42 & 3.28 & 2.86 & 3.39 & 2.70 & 2.68 & 2.60 \cite{naph_tetr1,naph_tetr2}\\
Pyrene      & 3.52 & 3.86 & 3.78 & 4.32 & 4.17 & 4.07 & 4.47 & 3.17 & 3.16 & 3.38 \cite{pyrene}\\
Penta.   & 1.83 & 2.17 & 2.11 & 2.79 & 2.68 & 3.17 & 2.74 & 2.32 & 2.45 & 2.14 \cite{pent}\\
C$_{96}$H$_{24}$      & 1.44 & 1.82 & 1.74 & 2.53 & 2.39 & 2.81 & 2.69 & 1.90 & 2.01 & 2.00 \cite{c96}\\ \hline
\textbf{MAE}         & 0.25	&0.27	&0.23	&0.78	&0.64	&0.84	&0.62	&0.13	&0.12	&\\
\end{tabular}
\end{ruledtabular}
\caption{\label{table1} $S_1$ excitation energies (eV) of the Flav dye family (6 dyes) and  6 planar hydrocarbons calculated using different TDDFT functionals, the individually fitted $v_W$, and the parameterized $v_W$ against experimental data.\cite{Cosco2017_ange,Cosco2020_naturechem,naph_tetr1,naph_tetr2,anth,pyrene,pent,c96} TD-LDA, TD-CAM-LDA0, and TDHF@$v_W$ are performed with in-house grid-based codes while the others are done in ORCA 6.0 with a def2-TZVPPD basis.\cite{Neese2020,Neese2022} The mean-absolute errors (MAE) reference to the experiments are shown for both sets.}
\label{tab:icg_hc}
\end{table*}

\begin{figure*}
\centering
\includegraphics[width=6.5in]{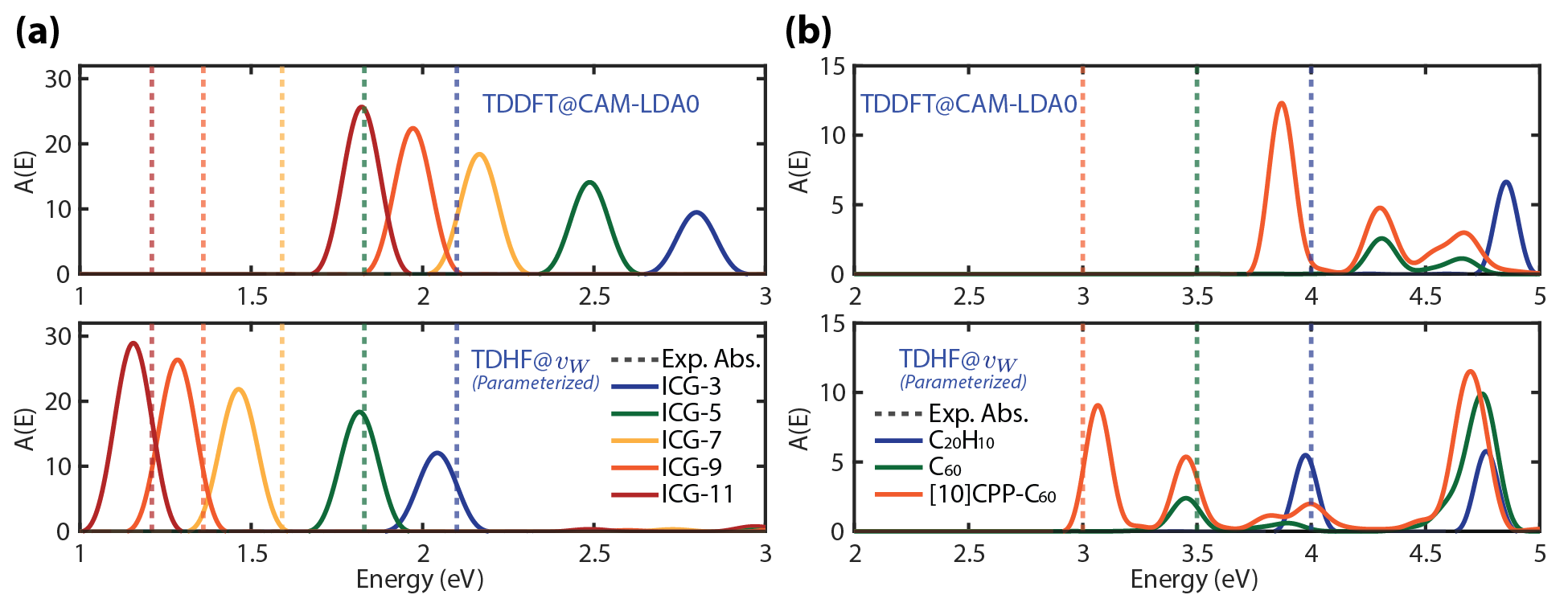}
\caption{\label{fig:fig_4} Linear response absorption spectra using TDDFT@CAM-LDA0 (upper panel), and TDHF@$v_W$ with parameterized $v_W$ (lower panel) for \textbf{(a)} ICG dye family and \textbf{(b)} curved hydrocarbons (C$_{20}$H$_{10}$, C$_{60}$, and [10]CPP-C$_{60}$). Experimental excitation peaks are shown in dotted lines, ICG data are from Refs. \cite{Swamy2024,Langhals2011,Gamage2024,Heng2022}, and curved hydrocarbons are from Refs. \cite{c20,c60,cpp10}.}
\end{figure*}

We also show in Fig. \ref{fig:fig_3} the $\epsilon^{-1}(k)$ along the $\hat{x}$, $\hat{y}$, and $\hat{z}$ directions overlaid by the isotropic-averaged fitting result. TDHF@$v_W$ spectra via the individually fitted and the parameterized $v_W$ are plotted together with the experimental $S_1$ peaks. The functional model with a $4^{\text{th}}$-order polynomial and an error function tail fits well in all 3 cases. The real-space representations of the inverse dielectric function, $\epsilon^{-1}(|r-r'|)$, for 3 molecules are shown in Appendix D. It is clear to see that the mid-range ($0.5-2.5\;\text{\AA}$) screening is significant for all of them, where the $\hat{z}$-direction has a stronger screening effect than $\hat{x}$ and $\hat{y}$ because there are no adjacent atoms in the $\hat{z}$-direction. Thus, electrons are more easily polarized by an external field along this direction. The $S_1$ peaks calculated for the three reference molecules from the parameterized $v_W$ match the ones obtained from the individually fitted $v_W$ to an accuracy of 0.01 eV (Fig. \ref{fig:fig_3}). The resulting parameters are tabulated in Table \ref{tab:functional_params}.

Table \ref{tab:icg_hc} shows the $S_1$ peak calculated from various TDDFT and TDHF methods for 2 families of molecules (6 Flav dyes and 6 planar hydrocarbons). Pure LDA has no explicit exchange, and bare HF has no electron correlation. Common hybrid-exchange functionals do not carefully treat the mid-range screening in the exchange.\cite{Jacquemin2007} Therefore, their mean-absolute errors (MAE) range between $\sim$ 0.3$-$1 eV for both families. MBPT-based methods, i.e., the $v_W$ approach significantly enhances the accuracy by pushing the MAE down to $\sim$ 0.1 eV or below. We show TDHF@$v_W$ results from both the individually fitted and parameterized. Note that we use the parameters obtained from Flav-9 for Flav dyes, and the parameters of pyrene for planar hydrocarbons. The parameterized $v_W$ gives slightly better MAE than the individually fitted one because the functional parameterization cancels out the stochastic noise. The resulting $\epsilon^{-1}(k)$ and the corresponding $v_W$ are smooth, which is a benefit of the present approach. We attribute the improved results for the parameterized $v_W$ versus the individually fit $v_W$ shown in Table \ref{tab:icg_hc} to the removal of the high-$k$ noise in the parameterized fit given by Eq. \ref{fit2}.

Fig. \ref{fig:fig_4} shows the $S_1$ peaks for (a) 5 ICG dyes and (b) 3 curved hydrocarbons: C$_{20}$H$_{10}$, C$_{60}$, and [10]CPP-C$_{60}$ against the experimental data. The upper panel illustrates the TDDFT results via CAM-LDA0 functional, and the lower panel shows the TDHF@$v_W$ (with parameterized $v_W$) spectra. The Flav-9 parameters are used again for ICG dye calculations, and the corannulene parameters are used for curved hydrocarbons. As expected, the more sophisticated treatment of the screening in the exchange kernel with $v_W$ enables accurate prediction of the optical gaps. Therefore, it produces absorption peaks that are close to the experimental references (MAE $<0.1$ eV).

\begin{figure}[H]
\centering
\includegraphics[width=3.5in]{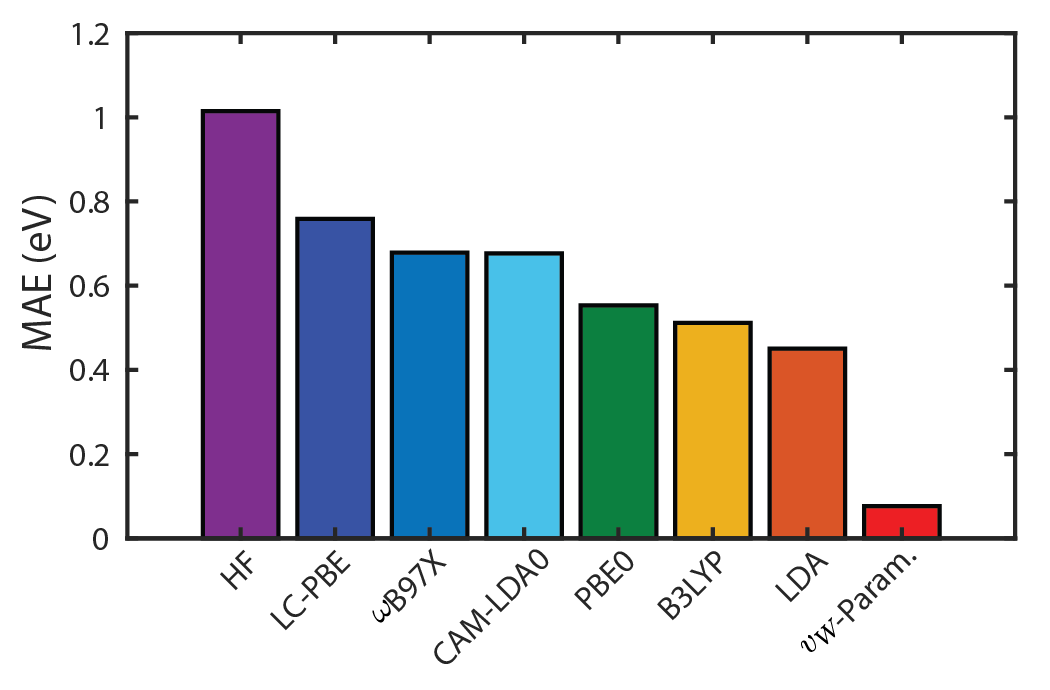}
\caption{\label{fig:fig_5} {MAE (in eV) of the optical gaps for 20 test molecules from various 
TDDFT/TDHF calculations against the experimental references.
}}
\end{figure}

{Fig. \ref{fig:fig_5} shows the MAE of} the optical gaps for all 20 systems studied in this work from various TDDFT{/TDHF methods} methods with the experimental references. The TDHF@$v_W$ (with parameterized $v_W$) achieves an MAE of 0.08 eV overall. In contrast, among all TDDFT functionals, TD-LDA produces an MAE of 0.45 eV,	TD-PBE0: 0.55 eV, TD-B3LYP: 0.51 eV, TD-LC-PBE: 0.76 eV, TD-$\omega$B97X: 0.68 eV, TD-CAM-LDA0: 0.68 eV, and TDHF: 1.01 eV. 

More importantly, since one set of parameters works for a family of molecules, and the parameters are grid-independent. The TDHF@$v_W$ calculation here scales essentially the same as a typical TDDFT but with an updated and more sophisticated exchange kernel.
\section{Conclusion}  
We introduce a parameterization scheme for the screened exchange kernel, $v_W$, which leverages the {photophysical} similarity among molecules within a given family (i.e., molecules that carry similar static dielectric responses) to bypass calculating and individually fitting the effective screened interaction, $W(r,r')$, for each molecule. Our approach provides a transferable, grid-independent kernel (with only 7 parameters) derived from a compact parameterization of the inverse dielectric function, $\epsilon^{-1}$. While chosen for mathematical convenience, our parameterization correctly reproduces on-site and asymptotic screening, ensuring physical fidelity. This method not only reduces computational overhead but also maintains accuracy in our TDHF@$v_W$ spectral calculations. 

We test this methodology across a diverse set of molecular systems, including $\pi$-conjugated polymethine cyanine dyes as well as planar and curved hydrocarbons. In each case, the $S_1$ excitation spectra show good agreement to results obtained from the individually fitted $v_W$ while also aligning well with experimental reference values, yielding an MAE $\sim 0.1$ eV. This consistency highlights the robustness and transferability of our parameterized kernel approach. Additionally, computational costs remain comparable to conventional TDHF or TDDFT with global or range-separated hybrid functionals while providing an improved treatment of electron-hole screening. 

While we have demonstrated this approach with three specific parameterization sets, it lays the foundation for broader applications, allowing the development of a standardized library of parameters that can facilitate efficient and scalable excitation spectra predictions for increasingly large and complex systems. Given its balance between computational efficiency and accuracy, MBPT-based TDHF@$v_W$ with a parameterized $v_W$ represents another step toward achieving BSE-quality spectral calculations at the computational cost of traditional TDHF, further bridging the gap between accuracy and feasibility in excited-state simulations. {As a future direction, self-consistently optimizing $v_W$ could further improve the accuracy of our approach, in the spirit of the bootstrap procedure for the exchange-correlation kernel developed by Gross et al.\cite{Sharma2011}.}

\section*{Appendix A: Hydrocarbon Structures}
\begin{figure}[H]
\centering
\includegraphics[width=3.1in]{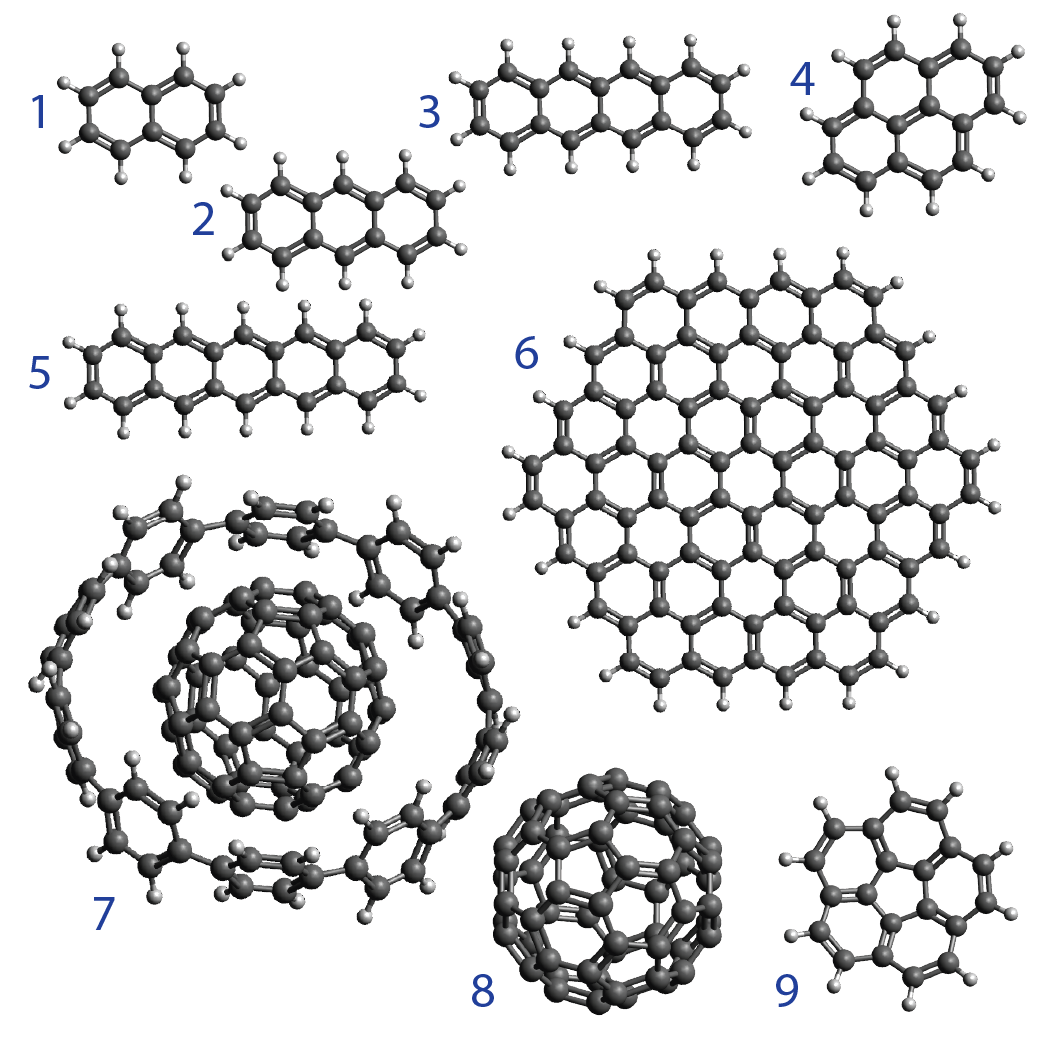}
\caption{\label{fig:hydrocarbons} Hydrocarbon systems tested in this work: (1) naphthalene (Naph.), (2) anthracene (Anth.), (3) tetracene (Tetra.), (4) pyrene, (5) pentacene (Penta.), (6) C$_{96}$H$_{24}$, (7) $[10]$CPP+C$_{60}$, (8) C$_{60}$, and (9) corannulene (C$_{20}$H$_{10}$).}
\end{figure}

{\section*{Appendix B: $k_{\text{mt}}$ Sensitivity Test}}
\begin{figure}[H]
\centering
\includegraphics[width=3.2in]{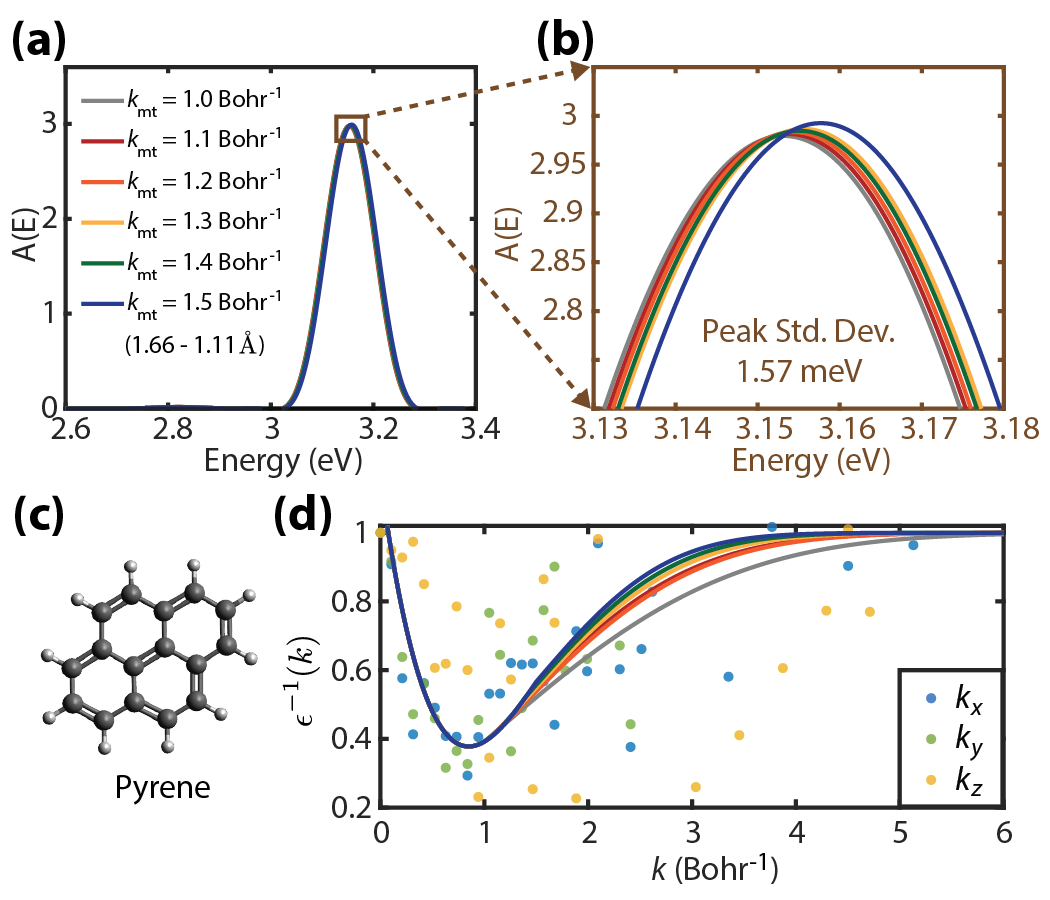}
\caption{\label{fig:kmt_conv} {\textbf{(a)} $S_1$ excitation peak of pyrene calculated via TDHF@$v_W$ with parameterized $v_W$ at different $k_{\text{mt}}$. \textbf{(b)} The zoomed-in peak positions of \textbf{(a)}, the standard deviation of the peak positions is 1.57 meV. \textbf{(c)} Molecular structure of pyrene. \textbf{(d)} $\epsilon^{-1}(k)$ parameterization with different $k_{\text{mt}}$, same color coding as in \textbf{(a)}.}}
\end{figure}

\section*{Appendix C: Computation Parameters}
\begin{table}[H]
\centering
\begin{ruledtabular}
\begin{tabular}{c|cccc|cc|c}
\textbf{System} & $n_x$ & $n_y$ & $n_z$ & $dr$ & $N_v$ & $N_c$ & $N_{k\text{-low}}$\\ \hline
ICG-3               & 110 & 100 & 80  & 0.4 & 105 & 400 & 953  \\
ICG-5               & 120 & 100 & 80  & 0.4 & 110 & 400 & 1045 \\
ICG-7               & 130 & 100 & 80  & 0.4 & 115 & 400 & 1127 \\
ICG-9               & 130 & 100 & 80  & 0.4 & 120 & 400 & 1127 \\
ICG-11              & 150 & 100 & 80  & 0.4 & 125 & 400 & 1293 \\ \hline
Flav-1              & 100 & 100 & 80  & 0.4 & 96  & 400 & 867  \\
Flav-3              & 110 & 100 & 80  & 0.4 & 101 & 400 & 953  \\
Flav-5              & 140 & 100 & 80  & 0.4 & 106 & 400 & 1195 \\
LFlav-7             & 140 & 100 & 80  & 0.4 & 111 & 400 & 1195 \\
Flav-7              & 140 & 100 & 80  & 0.4 & 122 & 400 & 1195 \\
Flav-9              & 120 & 120 & 120 & 0.5 & 116 & 400 & 3743 \\ \hline
C$_{20}$H$_{10}$    & 60  & 60  & 60  & 0.5 & 45  & 185 & 461  \\
C$_{60}$            & 60  & 60  & 60  & 0.5 & 120 & 400 & 461  \\
$[10]$CPP+C$_{60}$  & 80  & 80  & 50  & 0.5 & 260 & 800 & 1551 \\ \hline
Naph.               & 50  & 40  & 30  & 0.5 & 24  & 100 & 461  \\
Anth.               & 60  & 40  & 30  & 0.5 & 33  & 100 & 739  \\
Tetra.              & 70  & 40  & 30  & 0.5 & 42  & 150 & 1045 \\
Pyrene              & 60  & 60  & 60  & 0.5 & 37  & 150 & 461  \\
Penta.              & 80  & 40  & 30  & 0.5 & 51  & 200 & 1045 \\
C$_{96}$H$_{24}$    & 100 & 100 & 30  & 0.5 & 204 & 650 & 2801 \\
\end{tabular}
\end{ruledtabular}
\caption{Computational grids ($dr=dx=dy=dz$, unit: Bohr), $N_v$, and $N_c$. $N_{k\text{-low}}$: the number of deterministically treated long-wavelength terms, the high-$k$ space is represented with 1000 sparse stochastic vectors, details in Refs. \cite{bradbury_neargap_2023,Sereda2024}.}
\label{tab:grid_params}
\end{table}

\section*{Appendix D: Real-space Representation of $\epsilon^{-1}(|r-r'|)$}
\begin{figure}[H]
\centering
\includegraphics[width=3.5 in]{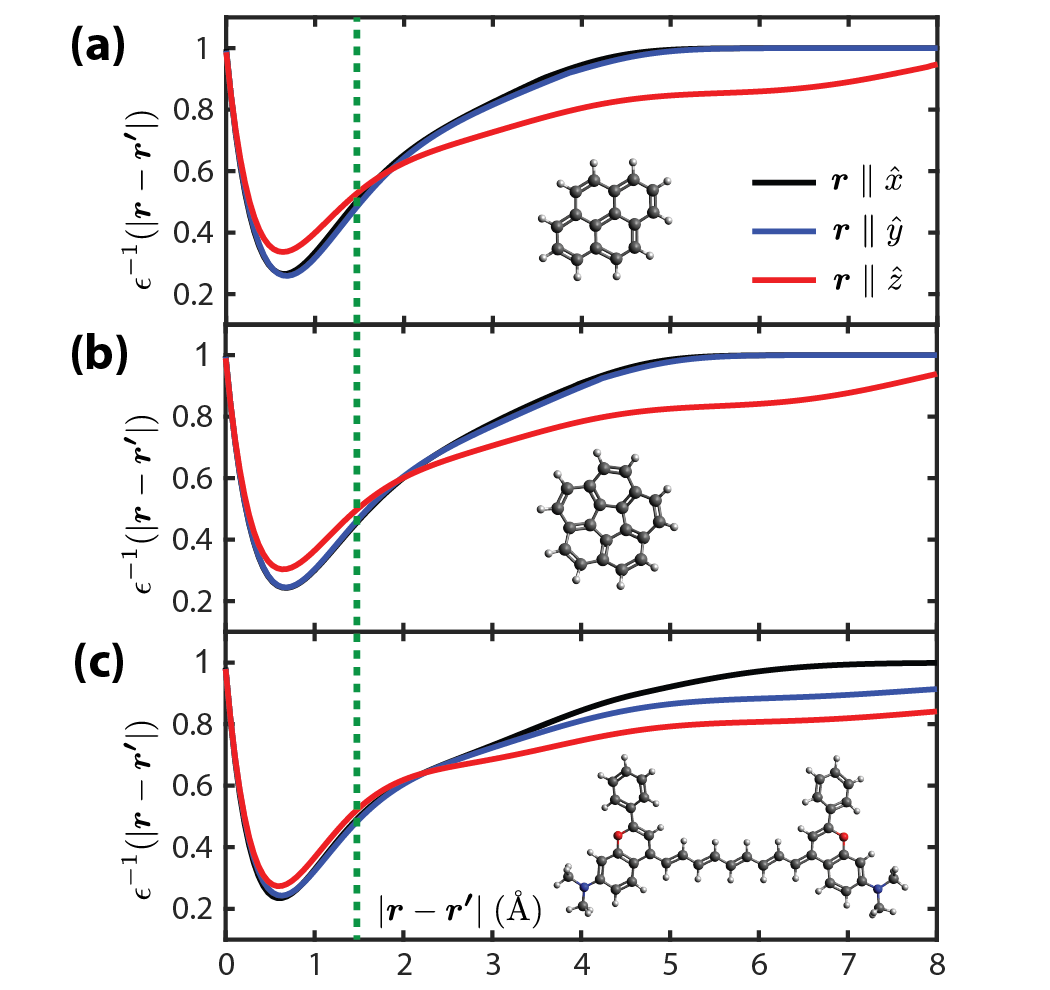}
\caption{\label{fig:vwr} Real-space inverse dielectric functions for \textbf{(a)} pyrene, \textbf{(b)} corannulene, \textbf{(c)} Flav-9. Green dotted line indicates the typical C=C bond length ($\sim1.4\;\text{\AA}$) in conjugated systems.}
\end{figure}

\begin{acknowledgments}
    We would like to express our deep appreciation to Prof. Abraham Nitzan for his enduring and foundational contributions to the field of theoretical physical chemistry. His insight and dedication have profoundly shaped the scientific community and continue to inspire our work. We also acknowledge the computational resources provided by the Expanse cluster at San Diego Supercomputer Center (SDSC) through allocation CHE-240183 under the Advanced Cyberinfrastructure Coordination Ecosystem: Services \& Support (ACCESS) program and the Hoffman2 Shared Cluster provided by UCLA Office of Advanced Research Computing’s Research Technology Group. JRC is supported by NSF grant CHE-2204263. DN is supported by NSF grant CHE-2245253.
\end{acknowledgments}

\section*{Data Availability}
The data that supports the findings of this study are available within the article and appendices. Additional data, source codes, {and tutorial videos} that support the findings {and reproducibilities} of this study are available from the corresponding author upon reasonable request.

\vspace{12cm}

\bibliography{apssamp}% Produces the bibliography via BibTeX.

\end{document}